\documentclass{aip-cp}
\usepackage[numbers,sort&compress]{natbib}
\usepackage{rotating}
\usepackage{graphicx}
\usepackage{indentfirst}
\usepackage{changepage}
\usepackage{lipsum}
\graphicspath{{./Figures/}}
\usepackage{savesym}
\savesymbol{iint}
\savesymbol{iiint}
\savesymbol{iiiint}
\savesymbol{idotsint}
\usepackage{amsmath}
\restoresymbol{AMS}{iint}
\restoresymbol{AMS}{iiint}
\restoresymbol{AMS}{iiiint}
\restoresymbol{AMS}{idotsint}
\usepackage{multirow}
\usepackage{xfrac}
\usepackage{amsfonts}

\usepackage{tikz}
\usepackage{lipsum}

\newcommand\copyrighttext{%
  \footnotesize The following article has been accepted by International Conference on Mechanical Engineering 2019 (ICME2019). After it is published, it will be found at https://publishing.aip.org/resources/librarians/products/journals/.}
\newcommand\copyrightnotice{%
\begin{tikzpicture}[remember picture,overlay]
\node[anchor=north,yshift=-30pt] at (current page.north) {\fbox{\parbox{\dimexpr\textwidth-\fboxsep-\fboxrule\relax}{\copyrighttext}}};
\end{tikzpicture}%
}

\begin{document}

\title{GPU Accelerated Lattice Boltzmann Simulation of Non-Newtonian Power-Law Fluid in a Porous Enclosure}

\author[aff1]{Mashnoon Islam}
\author[aff2,aff3]{Preetom Nag}
\author[aff2,aff3]{Md. Mamun Molla~\corref{cor1}}

\affil[aff1]{Department of Electrical and Computer Engineering, North South University, Dhaka, Bangladesh}
\affil[aff2]{Department of Mathematics $\&$ Physics, North South University, Dhaka-1229, Bangladesh}
\affil[aff3]{Center for Applied Scientific Computing (CASC), North South University, Dhaka-1229, Bangladesh}

\corresp[cor1]{Corresponding author: mamun.molla@northsouth.edu}

\maketitle
\copyrightnotice
\begin{adjustwidth}{3.81mm}{3.81mm}

\begin{abstract}
This paper demonstrates a numerical study of heat transfer in a square porous cavity filled with non-Newtonian power-law fluid. A Graphics Processing Unit (GPU) has been used to accelerate the numerical simulation, which uses the Multiple-Relaxation-Time (MRT) Lattice Boltzmann Method. A modified power-law model has been employed to characterize the flow of non-Newtonian fluids. The simulations have been conducted for the power-law index $n$ ranging from $(0.6 \leq n \leq 1.0)$, the Darcy number $Da$ ranging from $(10^{-3} \leq Da \leq 10^{-1})$ and the Rayleigh number $Ra$ ranging from $(10^3 \leq Ra \leq 10^5)$. Results show that the average Nusselt number ($\overline{Nu}$) decreases with an increase in the value of $n$ while $\overline{Nu}$ increases with an increase in the value of $Da$. Moreover, an increment in the value of $Ra$ leads to an increase in the average Nusselt number.
\end{abstract}
\end{adjustwidth}

\section{INTRODUCTION}\label{Intro:Natural Convection}
Convective heat transfer through porous media has received keen attention to the computational fluid dynamics(CFD) community because of its applications in different fields of engineering and science. Examples include civil engineering, mechanical engineering, chemical engineering, petroleum engineering, thermal management of electronic cooling, the improvement of heat transfer systems, and to name a few ~\cite{guo}. Related studies include simulation of natural convection of Newtonian and power-law fluids with laminar flow in a two-dimensional square enclosure with differentially heated sidewalls~\citet{turan11}. They have found that the mean Nusselt number ($\overline{Nu}$) increases with a higher Rayleigh number ($Ra$). However, the effect of higher Prandtl number to $\overline{Nu}$ is not prominent for both Newtonian and non-Newtonian fluids. They have also observed that $\overline{Nu}$ is higher for shear-thinning fluids ($n<1$) and has dropped down to 1 for shear thickening fluids($n>1$). \citet{turan12} have simulated the natural convection of power-law fluids with laminar flow in a two-dimensional square enclosed area with sidewalls differentially heated and subjected to constant heat flux. They have reported that $\overline{Nu}$ is higher for a higher $Ra$ for both fluids in both constant wall temperature configuration (CWT) and constant wall heat flux (CWHF) configuration. However, $\overline{Nu}$ is found to be lesser for CWHF than for the CWT condition. ~\citet{zhen} have simulated flow in a lid-driven cavity by Multiple-Relaxation-Time Lattice Boltzmann Method (MRT-LBM) for higher values of Reynolds numbers (20,000 $\le$ Re $\le 100,000$), and this study is the first attempt to simulate 2D cavity flow for a maximum of $Re=100,000$. Using MRT-LBM, fluid flow inside the two lid-driven cavities has been investigated by~\citet{guo2}. ~\citet{du} have investigated natural convection in a side-heated cavity using MRT-LBM to simulate fluid flow and SRT-LBM to simulate the temperature field. ~\citet{turan13} have simulated natural convection of power-law and Newtonian fluids with laminar flow in a two-dimensional rectangular enclosure where the sidewalls were differentially heated subject to both CWHF and CWT. They have found the unsteady and steady variation of $\overline{Nu}$  for both power-law fluids in the CWT condition and the CWHF condition, respectively.

Reports other than the above studies exist for non-Newtonian power-law fluids in porous media. However, some of those existing work do not agree with most of the similar validated work that has been done over the years. This study focus on a square porous cavity filled with non-Newtonian power-law fluids. The MRT-LBM simulation for this study has been conducted using a TESLA K40 GPU and the CUDA C platform.   

\section{MATHEMATICAL FORMULATIONS}\label{model:eps}
\begin{figure}[h]
  \centering
  \includegraphics[width=0.5\textwidth]{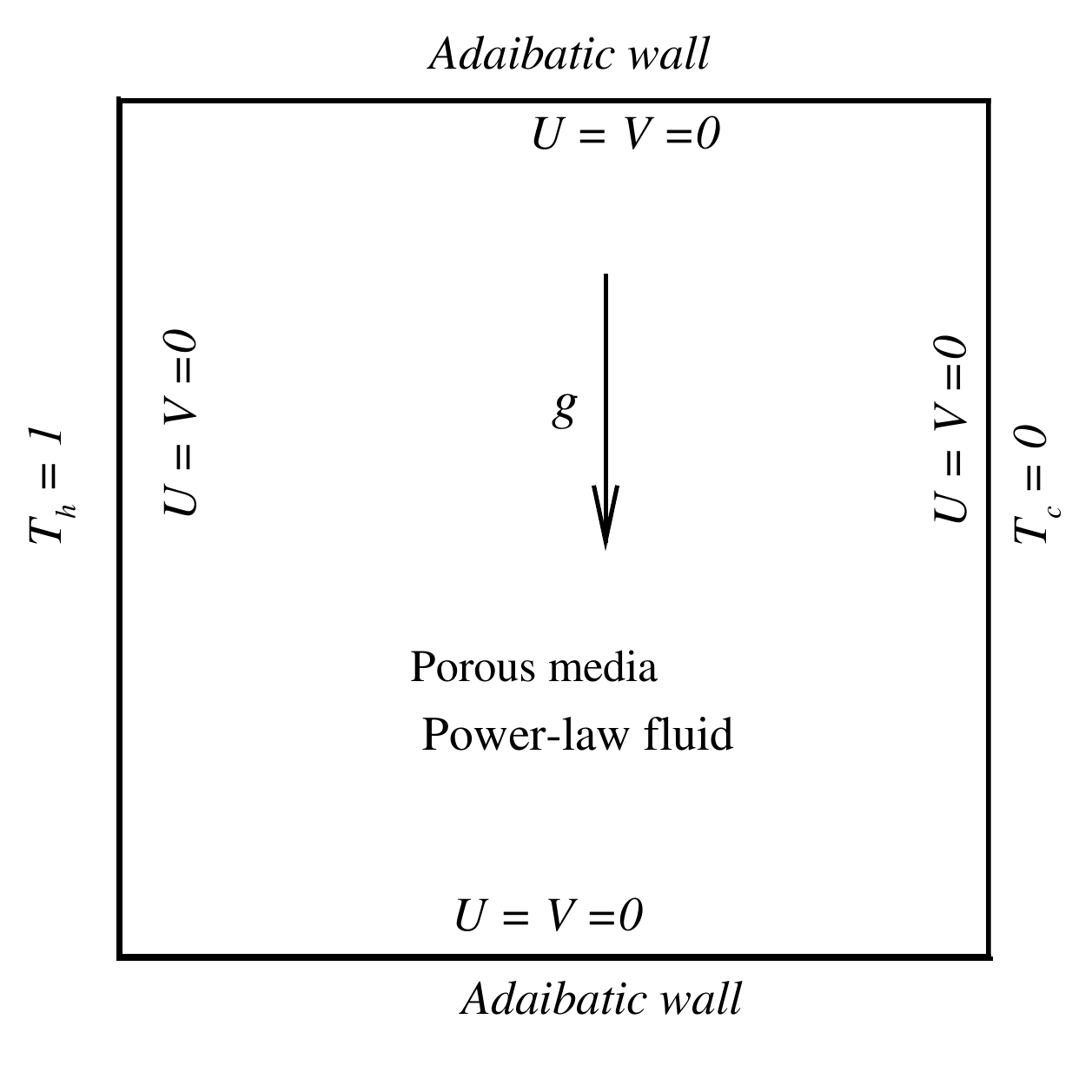}
  \caption{Simulation Geometry}
  \label{geometry}
\end{figure}

The geometry of the present study is considered as a square porous enclosure with height and width equal to $L$. The porous enclosure is filled with non-Newtonian power-law fluid. The left wall of the enclosure is maintained by a high temperature of $T_h$, which is the hottest part of the geometry. The rightmost wall has a temperature of $T_c$, which is the coldest part of the enclosure. The geometry is visualized in Figure \ref{geometry}.
The two-dimensional governing equations for conservation of mass, momentum and energy for non-Newtonian power-law fluids within porous enclosure can be written in the following dimensional form using the Boussinesq’s approximation:
\begin{equation}\label{continuity}
  \frac{\partial \bar{u}}{\partial \bar{x}} + \frac{\partial \bar{v}}{\partial \bar{y}}=0 
\end{equation}

\begin{equation}\label{u:momentum}
  \rho\left(\frac{\partial \bar{u}}{\partial \bar{t}} + \bar{u}\frac{\partial \bar{u}}{\partial \bar{x}} +\bar{v}\frac{\partial \bar{u}}{\partial \bar{y}}\right)=-\frac{\partial \bar{P}}{\partial \bar{x}} +  \frac{\partial}{\partial \bar{x}}\left(2\mu \frac{\partial \bar{u}}{\partial \bar{x}}\right) + \frac{\partial}{\partial \bar{y}}\left(\mu \frac{\partial \bar{u}}{\partial \bar{y}}\right) + \frac{\partial}{\partial \bar{y}}\left(\mu \frac{\partial \bar{v}}{\partial \bar{x}}\right) - \frac{\mu}{\kappa} \bar{u}
\end{equation}

\begin{equation}\label{v:momentum}
  \rho\left(\frac{\partial \bar{v}}{\partial \bar{t}} + \bar{u}\frac{\partial \bar{v}}{\partial \bar{x}} +\bar{v}\frac{\partial \bar{v}}{\partial \bar{y}}\right)=-\frac{\partial \bar{P}}{\partial \bar{y}} +  \frac{\partial}{\partial \bar{x}}\left(\mu \frac{\partial \bar{v}}{\partial \bar{x}}\right) + \frac{\partial}{\partial \bar{y}}\left(2\mu \frac{\partial \bar{v}}{\partial \bar{y}}\right) + \frac{\partial}{\partial \bar{y}}\left(\mu \frac{\partial \bar{u}}{\partial \bar{y}}\right) + g\rho \beta \left(T - T_C \right) - \frac{\mu}{\kappa} \bar{v} 
\end{equation}

\begin{equation}\label{energy}
  \frac{\partial \bar{T}}{\partial \bar{t}} + \bar{u}\frac{\partial \bar{T}}{\partial \bar{x}} +\bar{v}\frac{\partial \bar{T}}{\partial \bar{y}}=\alpha \left(\frac{\partial^2 \bar{T}}{\partial \bar{x}^2} + \frac{\partial^2 \bar{T}}{\partial \bar{y}^2} \right)
\end{equation}
The boundary conditions associated with the present study are as follows:
\begin{subequations}
  \begin{align}
    & \bar{u}=\bar{v}=0, T=T_H \quad  \text{on} \quad \bar{y}=0   
    & \bar{u}=\bar{v}=0, T=T_C \quad \text{on} \quad \bar{y}=H \\\
    & \bar{u}=\bar{v}=0, \frac{\partial \bar{T}}{\partial \bar{y}}=0 \quad \text{on} \quad \bar{x}=0  
    & \bar{u}=\bar{v}=0, \frac{\partial \bar{T}}{\partial \bar{y}}=0 \quad \text{on} \quad \bar{x}=L
  \end{align}
\end{subequations}
Here we introduce a set of non-dimensional variables to reduce the above dimensional equations Eq. \ref{continuity}-\ref{energy} into a non-dimensional form:
\begin{equation}\label{non_dim_var}
  \begin{split}
    & X=\frac{\bar{x}}{L}, \quad Y=\frac{\bar{y}}{L}, \quad U=\frac{\bar{u}}{\left( \frac{\alpha}{L} \right) \sqrt{Ra}}, \quad V=\frac{\bar{v}}{\left( \frac{\alpha}{L} \right) \sqrt{Ra}}, \quad Ra=\frac{g\beta L^{2n+1} (T_H-T_C)}{\nu \alpha^{n}} \\
    &\quad P=\frac{\bar{P}}{ \rho \left( \frac{\alpha}{L} \right)^{2} Ra}, \quad \tau=\frac{\bar{t} \sqrt{Ra}}{ \left( \frac {L^{2}}{\alpha} \right)}, \quad \theta=\frac{\bar{T} - T_C}{ T_H - T_C}, \quad Pr=\frac{\nu}{\alpha^{2-n}L^{2(n-1)}}, \quad Da=\frac{K}{L^2}
  \end{split}
\end{equation}
The non-dimensional form of the governing equations of the flow are obtained by:
\begin{equation}\label{par_sim_continuity}
  \frac{\partial U}{\partial X} + \frac{\partial V}{\partial Y}=0 
\end{equation}

\begin{equation}\label{par_sim_u:momentum}
  \frac{\partial U}{\partial \tau} + U \frac{\partial U}{\partial X} + V \frac{\partial U}{\partial Y}=- \frac{\partial P}{\partial X} + \frac{Pr}{\left( \sqrt{Ra}\right)^{2-n}} \left\{ \frac{\partial }{\partial X} \left(2D \frac{\partial U}{\partial X} \right) + \frac{\partial }{\partial Y} \left(D \frac{\partial U}{\partial Y} \right) + \frac{\partial }{\partial Y} \left(D \frac{\partial V}{\partial X} \right) - \frac{UD}{Da} \right\}
\end{equation}

\begin{equation}\label{par_sim_v:momentum}
  \frac{\partial V}{\partial \tau} + U \frac{\partial V}{\partial X} + V \frac{\partial V}{\partial Y}=- \frac{\partial P}{\partial Y} + \frac{Pr}{\left( \sqrt{Ra}\right)^{2-n}} \left\{ \frac{\partial }{\partial X} \left(D \frac{\partial V}{\partial X} \right) + \frac{\partial }{\partial Y} \left(2D \frac{\partial V}{\partial Y} \right) + \frac{\partial }{\partial Y} \left(D \frac{\partial U}{\partial Y} \right) - \frac{VD}{Da} \right\} + Pr \theta
\end{equation}

\begin{equation}\label{par_sim_energy}
  \frac{\partial \theta}{\partial \tau} + U \frac{\partial \theta}{\partial X} + V \frac{\partial \theta}{\partial Y}=\frac{1}{\sqrt{Ra}} \left(\frac{\partial^2 \theta}{\partial X^{2}} + \frac{\partial^2 \theta}{\partial Y^{2}} \right)
\end{equation}
where
\begin{equation}\label{D}
  D = \left\{ 2\left( \frac{\partial U}{\partial X} \right)^2 + 2\left( \frac{\partial U}{\partial Y} \right)^2 + \left( \frac{\partial U}{\partial Y} + \frac{\partial V}{\partial X} \right)^2  \right\}^{\frac{1}{2} \left( n - 1 \right)}
\end{equation}
where $U$ and $V$ correspond to the non-dimensional velocities of the non-Newtonian fluid along the non-dimensional $X$ and $Y$ directions. Other dimensionless quantities are pressure $P$ and temperature $\theta$ of the fluid. In Eq.~\ref{D}, $n$ is called the power-law index. For $n=1$, the fluid inside the porous enclosure behaves like a Newtonian fluid with an effective viscosity $\nu_0$. Therefore the deviation of $n$ from unity indicates the degree of deviation from  the Newtonian behavior. Note that the case $n < 1.0$ corresponds to the shear-thinning (pseudo-plastic) fluid, whereas $n > 1.0$ denotes the shear-thickening (dilatant) fluid.

\subsection{Multiple-Relaxation-Time Lattice Boltzmann Method}\label{normalMRT}
The D2Q9 model of the Lattice Boltzmann Method has been used for simulating the velocity field. The related equations have been discussed below.

The Multiple-Relaxation-Time Lattice Boltzmann equation with collision operation is as follows:
\begin{equation}\label{normalMRTequation}
  f( x + e_i \Delta t, t + \Delta t) - f( x, t) = -M^{-1} S \left\{ m(x, t) - m^{eq}(x, t) \right\} + M^{-1} \left( I - \frac{S}{2} \right) F( x, t)
\end{equation}
where the distribution function is $f = (f_0, f_1, f_2, f_3...f_n)^{T}$, the vector of moments are $m = (m_0, m_1, m_2...m_n)^{T}$ and $m^{eq} = (m_0^{eq}, m_1^{eq}, m_2^{eq}...m_n^{eq})^{T}$, and the force components are $F = (F_0, F_1, F_2, F_3...F_n)^{T}$. Here, the diagonal collision matrix $S$ is defined by:
\begin{equation}\label{collisionMatrix}
S=diag[s_0, s_1, s_2, s_3, s_4, s_5, s_6, s_7, s_8]
\end{equation}
where $s_0 = s_3 = s_5 = 1.0$, $s_1 = s_2 = 1.4, s_4 = s_6 = 1.2$ and $s_7 = s_8 = \frac{1}{\tau}$ where $\tau = \frac{3v + 1}{2}$ is defined as the relaxation time. The particle speed $e_i$ is expressed as:
\begin{equation}\label{particleSpeedMatrix}
  e_i = 
  \begin{cases}
    (0, 0) & i = 0 \\\
    c(1, 0), c(0, 1), c(-1, 0), c(0, -1) & i = 1, 2, 3, 4 \\\
    c(1, 1), c(-1, 1), c(-1, -1), c(1, -1) & i = 5, 6, 7, 8 
  \end{cases}
\end{equation}
where $c^{def} = \frac{\Delta x}{\Delta t}$ is known as the lattice speed. The kinematic viscosity is: denoted by $\nu$ and is defined as follows:
\begin{equation}\label{kinematic_viscosity}
  \nu(\bar{x},t) = c_s^2 \Delta t \left(\tau - \frac{a}{2} \right), \quad c_s^2 = \frac{1}{3}
\end{equation} 

The macroscopic fluid density is denoted by $\rho$. The velocity is denoted as $u$ and both have been derived from the distribution function of the moments. The definitions are given in equations (\ref{macro_fluid_density}) and (\ref{velo}) respectively:
\begin{equation}\label{macro_fluid_density}
  \rho = \sum_{i=0}^{g} f_i
\end{equation}

\begin{equation}\label{velo}
u = \frac{1}{\rho} \sum_{i=0}^{g} f_i e_i + \frac{G}{2}
\end{equation}
where $G$ is the buoyancy term and is defined as $G=g\beta(T-T_{m})$.

\subsection{Multiple-Relaxation-Time Thermal Lattice Boltzmann Method}\label{thermalMRT}
The D2Q5 model of the Thermal Lattice Boltzmann Method has been used for simulating the temperature field. The Multiple-Relaxation-Time Thermal Lattice Boltzmann equation with collision operation is as follows:
\begin{equation}\label{thermalMRTequation}
  g( x + e_i \Delta t, t + \Delta t) - g( x, t) = -N^{-1} S \left\{ m(x, t) - m^{eq}(x, t) \right\} 
\end{equation}
where the thermal distribution function is $g = (g_0, g_1, g_2...g_n)^{T}$ and N is the 5 by 5 collision matrix. The discrete velocities are given by:
\begin{equation}\label{particleSpeedMatrix}
  e_i = 
  \begin{cases}
    (0, 0) & i = 0 \\\
    c(1, 0), c(0, 1), c(-1, 0), c(0, -1) & i = 1, 2, 3, 4
  \end{cases}
\end{equation}
The temperature is denoted by $T$ and is defined as:
\begin{equation}\label{particleSpeedMatrix}
e_i = T=\sum_{i=0}^{4} g_i
\end{equation}
The detailed formulation and boundary conditions are given in the previous paper of Molla et. al.~\cite{Molla_2018}.

\subsection{Power-law Viscosity Model}
In this paper, a singularity-free modified power-law model has been adopted~\cite{Thohura_2019} to quantify the non-Newtonian viscosity for a fluid.

\begin{equation}\label{Eq:MPL}  
  \nu(\bar{x},t)=\left\{\begin{array}{l} {\nu _0 },  \quad \quad \quad \quad \quad \quad \left(\left|{\gamma }\right|\right)< {\gamma }_{1} \\ 
  {\nu _{0}\left(\left|{\gamma }\right|\right)^{n-1},\quad \quad {\gamma }_{1} \le \left(\left|{\gamma }\right|\right)\le{\gamma }_{2} } \\ {\nu _0\left(\gamma_2\right)^{n-1} },  \quad \quad \quad \left(\left|{\gamma }\right|\right)> {\gamma }_{2} \end{array}\right.,
 \end{equation} 
where $\gamma_1 =0.01  $ and $\gamma_2 = 10^6 $  are  the lower and upper threshold shear-rate limits respectively. If the shear rate is less than the lower-threshold shear rate limit $\gamma_1$ and greater than the upper-threshold shear-rate limit  $\gamma_2$, the viscosity behaves like that of a Newtonian fluid. Here, $\nu_0=\frac{Pr}{\sqrt{Ra}^{2-n}}$ is chosen to be the constant kinematic viscosity for the Newtonian fluid. For the MRT model shear rate or strain rate tensor $\gamma_{\alpha\beta}$ can be calculated locally as:
 \begin{equation}
  \gamma_{\alpha\beta}= -\frac{3}{2\rho c^2 }\sum^{8}_{i= 0} {e}_{i\alpha}{e}_{i\beta}\sum^{8}_{j= 0}\left(\bf{M^{-1}\bf{S}\bf{M}}\right)_{ij}\left(f_j-f_{j}^{(eq)}\right),
\end{equation}
where, $f^{(eq)}$ is the equilibrium distribution function. The magnitude of the shear rate $\gamma_{\alpha\beta}$ is obtained using the Frobenius norm:
\begin{equation}
  |\gamma|=\sqrt{2\gamma_{\alpha\beta} \gamma_{\alpha\beta}}.
\end{equation}

\section{AVERAGE RATE OF HEAT TRANSFER}\label{avgHeatTransfer}
For the natural convection flow, the substantial interest as for the engineering quantity is the heat transfer rate from the hot (left) wall, which is determined by the local Nusselt number~$(Nu)$. The average Nusselt number can then be evaluated as:
\begin{equation}\label{avgNu}
  \overline{Nu}=\frac{1}{L} \int_{0}^{L} Nu(Y) dY, \quad \text{where} \quad  Nu(Y)=-\frac{\partial \theta}{\partial X}|_{X=0}
\end{equation}

\section{CODE VALIDATION}
In order to validate the present simulation code, a comparison is made between the present results and the benchmark results published by \citet{turan11} and \citet{matin13} for different power-law index ($n=0.6, 1.0$ and $1.4$). As shown in Table \ref{tab:val}, results using the present code are agree well with the benchmark results published before. Thus, it can be concluded that the present code is producing correct simulation data for the non-Newtonian power-law fluids. Validation of the same code for the Darcy case has been reported by \citet{Molla_2018} recently.
\begin{table}[h]
  \label{tab:val} 
  \caption{Comparison of present $\overline{Nu}$ with that of \citet{matin13} and \citet{turan11} for the non-Newtonian case ($Ra=10^5$ and $Pr=100$)} \\ 
  \centering
    \begin{tabular}{c c c c}
      \hline
      & & $\overline{Nu}$ & \\ \cline{2-4}
      & $n=0.6$ & $n=1.0$ & $n=1.4$ \\ 
      \hline
      Present & 12.95289 & 4.69148 & 2.44821 \\ 
      \hline 
      \citet{matin13} & 13.06722 & 4.69312 & 2.28356 \\
      \hline
      \citet{turan11} & 12.98500 & 4.72576 & 2.28945 \\ 
      \hline
  \end{tabular}
\end{table}

\section{RESULTS AND DISCUSSIONS}\label{results}
Before conducting the simulations, a grid independence test has been accomplished using three different grids of $128 \times 128$, $160 \times 160$, and $192 \times 192$  (tests are not shown here). Results show that there exists a very negligible effect on the results of the simulations. Therefore, the simulation is conducted using the grid size of $160\times160$ in order to compromise between high accuracy and computational efficiency.

The simulation procedure has been divided into three cases, spawning a total of 9 simulations. The results, along with the configurations of the simulations of Case 1, Case 2, and Case 3, are summarized in Table \ref{tab:one}.

\begin{figure}
  \centering
  \begin{tabular}[h]{c}
    \includegraphics[width=0.7\textwidth]{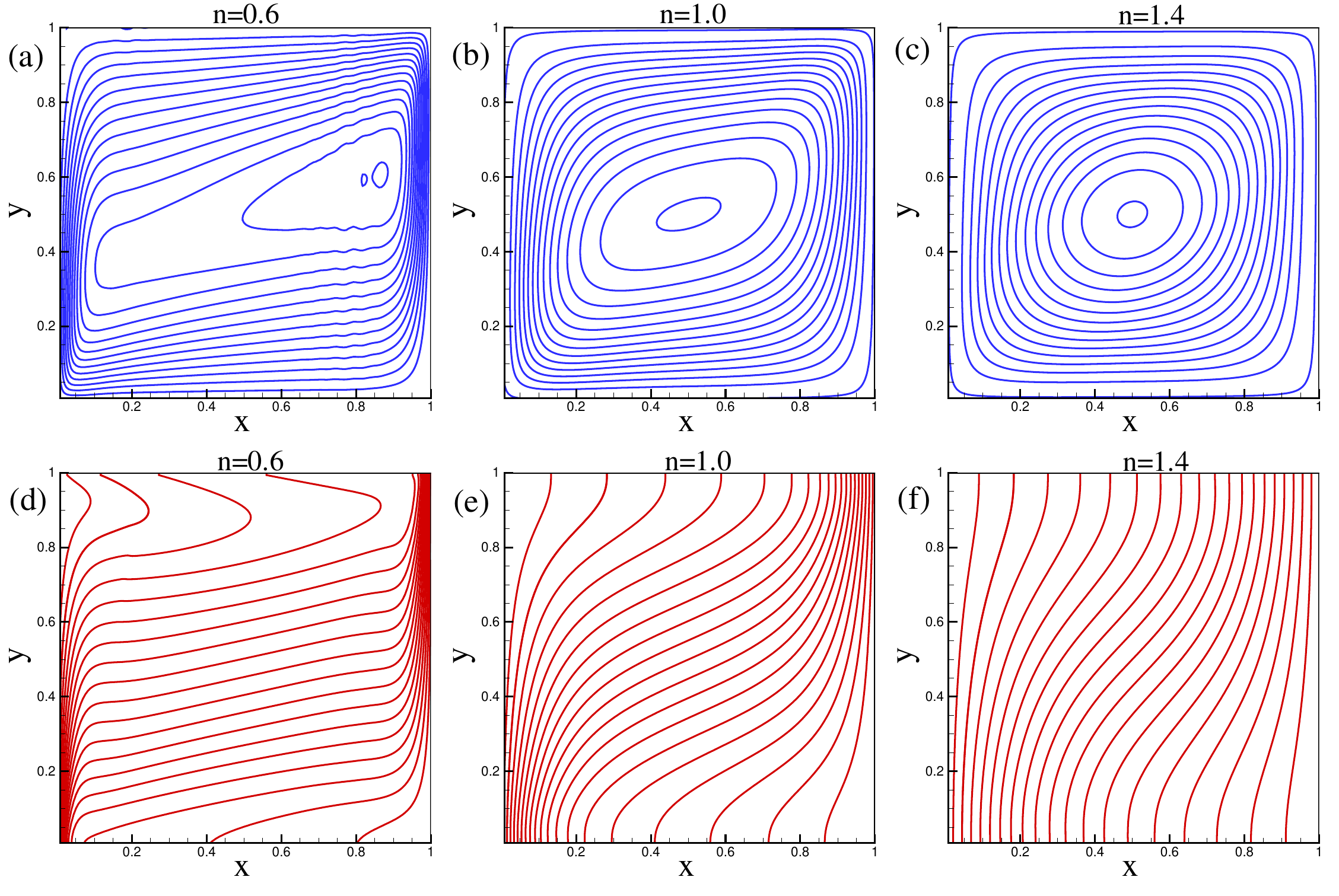}\label{fig:str_isotherms_case_1} \\
  \end{tabular} 
  \caption{Streamlines (a, b, c) and Isotherms (d, e, f) for Case 1: $Da=10^{-3}$, $Pr=6.2$, $Ra=10^5$, $0.6 \leq n \leq 1.4$}
\end{figure}

\begin{table}[h]
  \caption{Results' Summary of Case $1$, Case $2$ and Case $3$} \\
  \label{tab:one}
  \begin{tabular}{c c c}
    \begin{tabular}{c c}
      \multicolumn{2}{c}{Case 1} \\ 
      \multicolumn{2}{c}{$Da=10^{-3}$, $Pr=6.2$, $Ra=10^5$} \\  
      \hline
      $n$ & $\overline{Nu}$ \\ [0.5ex]
      \hline 
      0.6 & 9.249539 \\ 
      \hline 
      1.0 & 2.430419 \\
      \hline
      1.4 & 1.300101 \\ [0.5ex] 
      \hline
    \end{tabular}
    &
    \begin{tabular}{c c}
      \multicolumn{2}{c}{Case 2} \\  
      \multicolumn{2}{c}{$n=0.6$, $Pr=6.2$, $Ra=10^5$} \\ 
      \hline
      $Da$ & $\overline{Nu}$ \\ [0.5ex] 
      \hline
      $10^{-3}$ & $9.249539$ \\ 
      \hline 
      $10^{-2}$ & $12.835687$ \\
      \hline
      $10^{-1}$ & $13.316177$ \\ [0.5ex] 
      \hline
    \end{tabular}
    &
    \begin{tabular}{c c}
      \multicolumn{2}{c}{Case 3} \\ 
      \multicolumn{2}{c}{$n=0.6$, $Pr=6.2$, $Da=10^{-2}$} \\   
      \hline
      $Ra$ & $\overline{Nu}$ \\ [0.5ex] 
      \hline
      $10^{3}$ & $1.068697$ \\ 
      \hline 
      $10^{4}$ & $4.411946$ \\
      \hline
      $10^{5}$ & $12.834702$ \\ [0.5ex] 
      \hline
    \end{tabular}
  \end{tabular}
\end{table}
For Case 1, the power-law index ($n$) has been varied from $0.6$ to $1.4$ while the Darcy number ($Da$), Prandtl number ($Pr$) and Rayleigh number ($Ra$) have been kept fixed by $10^{-3}$, $6.2$ and $10^{5}$ respectively. The results for Case 1 in Table \ref{tab:one} show that the average Nusselt number ($\overline{Nu}$) decreases with an increase in $n$. The results indicate that the average rate of heat transfer from the hot surface to the enclosure becomes maximized for shear-thinning fluids ($n < 1$) and minimized for shear-thickening fluids ($n > 1$). The phenomenon is coupled with the distribution of velocity and temperature within the boundary layer near the hot wall. In Figure \ref{fig:str_isotherms_case_1}, it is observed that the streamlines are denser adjacent to the hot walls for shear-thinning fluids than other kinds of fluids. As a result, the boundary layer becomes thicker for shear-thickening fluids than the shear-thinning fluids. It happens because shear-thickening fluids generate higher viscosity than shear-thinning fluids. Noteworthy, the spaces between streamlines are narrower in Figure \ref{fig:str_isotherms_case_1}(a), compared to Figure \ref{fig:str_isotherms_case_1}(b) and (c), as $n$ is decreased from $1.4$ to $0.6$. The phenomenon indicates the slower velocity for shear thickening fluids. Also, the central vortex is shifting towards the right side with making an angle with the horizontal axis as one changes the fluids from shear thickening to shear-thinning fluid. 

From the results of the isotherms in Figure \ref{fig:str_isotherms_case_1}(d)-(f), it is observed that the thermal boundary layer becomes thinner for shear-thinning fluids. The phenomenon is a result of the high-temperature gradient and heat transmission from the hot wall to the enclosure. In Figure \ref{fig:str_isotherms_case_1}(f), less isotherms reach the cold wall and are less curved compared to Figure \ref{fig:str_isotherms_case_1}(d) and Figure \ref{fig:str_isotherms_case_1}(e), indicating poor convective performance, and thus, less heat transfer. Our findings agree with that of \cite{turan11, binkim}.

For the second case in Table \ref{tab:one}, $Da$ has been varied from $10^{-3}$ to $10^{-1}$ and other parameters like $n$, $Pr$ and $Ra$ have been kept fixed by $0.6$, $6.2$ and $10^{5}$, respectively. The results for the average Nusselt number has been tabulated in Table \ref{tab:one}. Results show that $\bar{Nu}$ increases with an increase of $Da$. Therefore, the average rate of heat transfer becomes maximized if the media is less porous and generates high permeability, i.e., higher values of $Da$.  Although results are not shown here, however, we observed that the streamlines near the center for $Da=10^{-3}$ become narrower than that for $Da=10^{-1}$. As a result, relatively poor convection and thus less heat transfer are obtained for lower values of $Da$. It is happened because of the enhancement of hydraulic resistance of the porous medium with decreasing the Darcy number $Da$. Thus, increasing $Da$ leads to an increase in the fluid flow in the porous medium and enhances the heat transfer rate inside the enclosure. The isotherms are found to be more horizontal for $Da=10^{-1}$ than that for $Da=10^{-3}$, indicating better heat transfer for less porous media. Other researchers also reported similar phenomena before~\cite{basak, yaghou}.

In case 3, $Ra$ has been varied from $10^{3}$ to $10^{5}$ . The results for case 3 in Table \ref{tab:one} show that $\bar{Nu}$ increases with an increase in $Ra$, which also agrees with conclusions of \cite{yaghou, turan11}. Thus, the heat transfer rate increases with increasing the buoyancy force. 

\section{CONCLUSION}
This paper demonstrates heat transfer through natural convection of non-Newtonian power-law fluids in a square porous enclosure. An empirical power-law model has been used to demonstrate the effects of shear-thinning and shear-thickening of a fluid. Among the three cases presented in this paper, we can conclude that:

\begin{itemize}
  \item The results of the first case show that the average heat transfer rate from the surface to the enclosure~$(\overline{Nu})$ decreases with an increase in $n$.  
  \item The results of the second case show that $\overline{Nu}$ increases with an increase of Darcy number~$(Da)$.
  \item Lastly, increasing the Rayleigh number ~$(Ra)$ leads to enhance the surface heat transfer rate inside the enclosure.
\end{itemize}

\section{ACKNOWLEDGMENTS}
This work is supported by Faculty Research Grant~(NSU-RP-18-067 \& CTRG-19/SEPS/15), North South University, Bashundhara, Dhaka 1229, Bangladesh. The third author would like to thank the NVIDIA Corporation, USA, for granting the TESLA k40 GPU for facilitating GPU computation. 

\bibliographystyle{aipnum-cp}
\bibliography{reference_cite}

\end{document}